\documentclass[manuscript]{aastex}
\usepackage{color,amssymb,amsmath,graphicx,longtable,lscape,natbib,multirow}

\newcommand{\nuclei}[2]{\ensuremath{\mathrm{^{#1}#2}}}

\newcommand{\helium}[1][4]{\nuclei{#1}{He}}

\newcommand{\carbon}[1][12]{\nuclei{#1}{C}}

\newcommand{\oxygen}[1][16]{\nuclei{#1}{O}}

\newcommand{\silicon}[1][28]{\nuclei{#1}{Si}}

\newcommand{\nickel}[1][58]{\nuclei{#1}{Ni}}

\begin{document}

\title{Zero Impact Parameter White Dwarf Collisions in FLASH}
    
\author{
W.P. Hawley\altaffilmark{1}, 
T. Athanassiadou\altaffilmark{1,2},
F.X. Timmes\altaffilmark{1,2}
        }
\altaffiltext{1}{School of Earth and Space Exploration,
                 Arizona State University,
                 Tempe, AZ 85287, USA}
\altaffiltext{2}{The Joint Institute for Nuclear Astrophysics,
		Notre Dame, IN 46556, USA}
\email{Wendy.Hawley@asu.edu}

\begin{abstract}
We systematically explore zero impact parameter collisions of white
dwarfs with the Eulerian adaptive grid code FLASH for 0.64+0.64
M$_{\odot}$ and 0.81+0.81 M$_{\odot}$ mass pairings. Our models span a
range of effective linear spatial resolutions from 5.2$\times$10$^{7}$
to 1.2$\times$10$^{7}$ cm.  However, even the highest resolution 
models do not quite achieve strict numerical convergence, due to the
challenge of properly resolving small-scale burning and energy
transport.  The lack of strict numerical convergence from these 
idealized configurations suggest that quantitative predictions of 
the ejected elemental abundances that are generated by binary white dwarf 
collision and merger simulations should be viewed with caution.
Nevertheless, the convergence trends do allow some patterns
to be discerned. We find that the 0.64+0.64 M$_{\odot}$ head-on collision model
produces 0.32 M$_{\odot}$ of \nickel[56] and 0.38 M$_{\odot}$ of
\silicon[28], while the 0.81+0.81 M$_{\odot}$ head-on collision model
produces 0.39 M$_{\odot}$ of \nickel[56] and 0.55 M$_{\odot}$ of
\silicon[28] at the highest spatial resolutions.  Both mass pairings
produce $\sim$0.2 M$_{\odot}$ of unburned \carbon[12]+\oxygen[16]. We
also find the 0.64+0.64 M$_{\odot}$ head-on collision begins 
carbon burning
in the central region of the stalled shock between the two white
dwarfs, while the more energetic \hbox{0.81+0.81 M$_{\odot}$} head-on
collision raises the initial post-shock temperature enough to burn the entire
stalled shock region to nuclear statistical equilibrium.
\end{abstract}

\section{Introduction}
Supernova Type Ia (SNIa) have continued to be foremost probes of the
universe's accelerating expansion
\citep{riess_1998_aa,perlmutter_1999_aa,riess_2011_aa,sullivan_2011_aa,
  suzuki_2012_aa}.  While light curves between different SNIa vary,
the variations generally correlate with distance independent
light-curve properties, such as the decline from B band maximum after
15 days \citep{phillips_1993_aa}. Calibration of the light curves onto
a standard template yields distance indicators accurate to $\sim$10\%
\citep[e.g.,][]{silverman_2012_aa} and are primarily applied to SNIa
not showing peculiarities \citep{branch_1993_aa}.  These ``normal''
SNIa presumably emerge from a homogeneous population of white dwarf
progenitors.  While the favored population is thought to be a
carbon-oxygen white dwarf (WD) accreting matter from a non-degenerate
companion star \citep[e.g.,][]{whelan_1973_aa, thielemann_1986_aa},
recent observations suggest that a fraction of SNIa may derive from
double-degenerate progenitors
\citep{howell_2006_aa,hicken_2007_aa,gilfanov_2010_aa,bianco_2011_aa}.

In view of these and other observations of SNIa progenitor systems,
recent theoretical studies have explored double degenerate mergers and
collisions of white dwarfs as potential progenitors of SNIa
\citep{guerrero_2004_aa, yoon_2007_aa, maoz_2008_aa,
  loren-aguilar_2009_aa, raskin_2009_aa, rosswog_2009_aa,
  loren-aguilar_2010_aa, pakmor_2010_aa, raskin_2010_aa, shen_2012_aa,
  pakmor_2012_aa}.  Almost all of these efforts use smooth particle
hydrodynamic (SPH) codes to model most of the collision or merger
process.  SPH and Eulerian grid codes, such as FLASH
\citep{fryxell_2000_aa}, have well-known complimentary strengths and
weaknesses $-$ particle codes are inherently better at angular
momentum conservation, whereas grid codes have a superior treatment of
shocks. Only \citet{rosswog_2009_aa} included a zero impact parameter
white dwarf collision model with FLASH. They used a mirror
gravitational potential for one white dwarf at one spatial
resolution. They found their FLASH calculations yielded about half as
much \nickel[56] as the equivalent SPH calculation (0.32 M$_{\odot}$
for SPH, 0.16 M$_{\odot}$ for FLASH).

In this paper, we use the Eulerian adaptive mesh refinement code FLASH
to model the zero impact collisions between 0.64+0.64 M$_{\odot}$ and
0.81+0.81 M$_{\odot}$ carbon-oxygen white dwarf mass pairings.  
Like the single case studied by \citet{rosswog_2009_aa}, our
configurations are highly idealized cases of head-on collisions
between identical, initially spherical white dwarfs. One aim of our
paper is to determine whether or not, given presently available
computing resources and numerical algorithms, simulations of
collisions can be used to reliably predict the fraction of 
white dwarf material that is converted by explosive nucleosynthesis into 
heavier elements such as silicon and nickel. 
Other efforts have focused on the realism of the initial conditions
and subsequent evolution, including but limited to, 
in-spiraling from a binary orbit
\citep{rasio_1995_aa,pakmor_2010_aa,dan_2011_aa,raskin_2012_aa}, 
unequal mass collisions \citep{benz_1989_aa,benz_1990_aa,rosswog_2009_aa,
loren-aguilar_2010_aa, raskin_2010_aa,pakmor_2012_aa},
and the final long-term fate of merged systems
\citep{van-kerkwijk_2010_aa,yoon_2007_aa,shen_2012_aa}.  Our
simulations, through their idealized nature, highlight the essential
physics and numerical convergence properties of the simplest possible
configuration. In addition, our idealized configurations form a
baseline for further studies that incorprate more realistic initial
conditions.

Our paper is organized as follows. In \S\ref{s.simulations}, we
describe the input physics, initial conditions, and boundary
conditions of our FLASH simulations.  In \S\ref{s.results}, we discuss
the results of our studies over a range of spatial resolutions and
time-step choices, and in \S\ref{s.discuss} we explore the
implications of our results and describe directions for future
studies.

\section{Input Physics, Initial Conditions, and Boundary Conditions}
\label{s.simulations}

Our 3D simulations are carried out with FLASH 3.2, a 3D Eulerian
hydrodynamics code with adaptive mesh refinement
\citep{fryxell_2000_aa, calder_2002_aa}. We use the included Helmholtz
equation of state \citep{timmes_2000_ab}, the 13 isotope
alpha-chain reaction network that includes isotopes from \helium[4] to
\nickel[56] to model energy generation from nuclear burning
\citep{timmes_1999_ab}, and the multigrid Poisson gravity
solver with Dirichlet boundaries \citep{ricker_2008_aa}.
All the simulation domain boundaries use a diode boundary condition, 
which is a zero-gradient boundary condition where fluid velocities are
not allowed to point back into the domain. 
We follow both
white dwarfs in 3D rectilinear coordinates throughout calculation, rather
than using a mirror gravitational potential and evolving one white dwarf
\citep[as used in][]{rosswog_2009_aa}.

\begin{figure}[tbp]
\begin{center}
\includegraphics[width=1.0\textwidth]{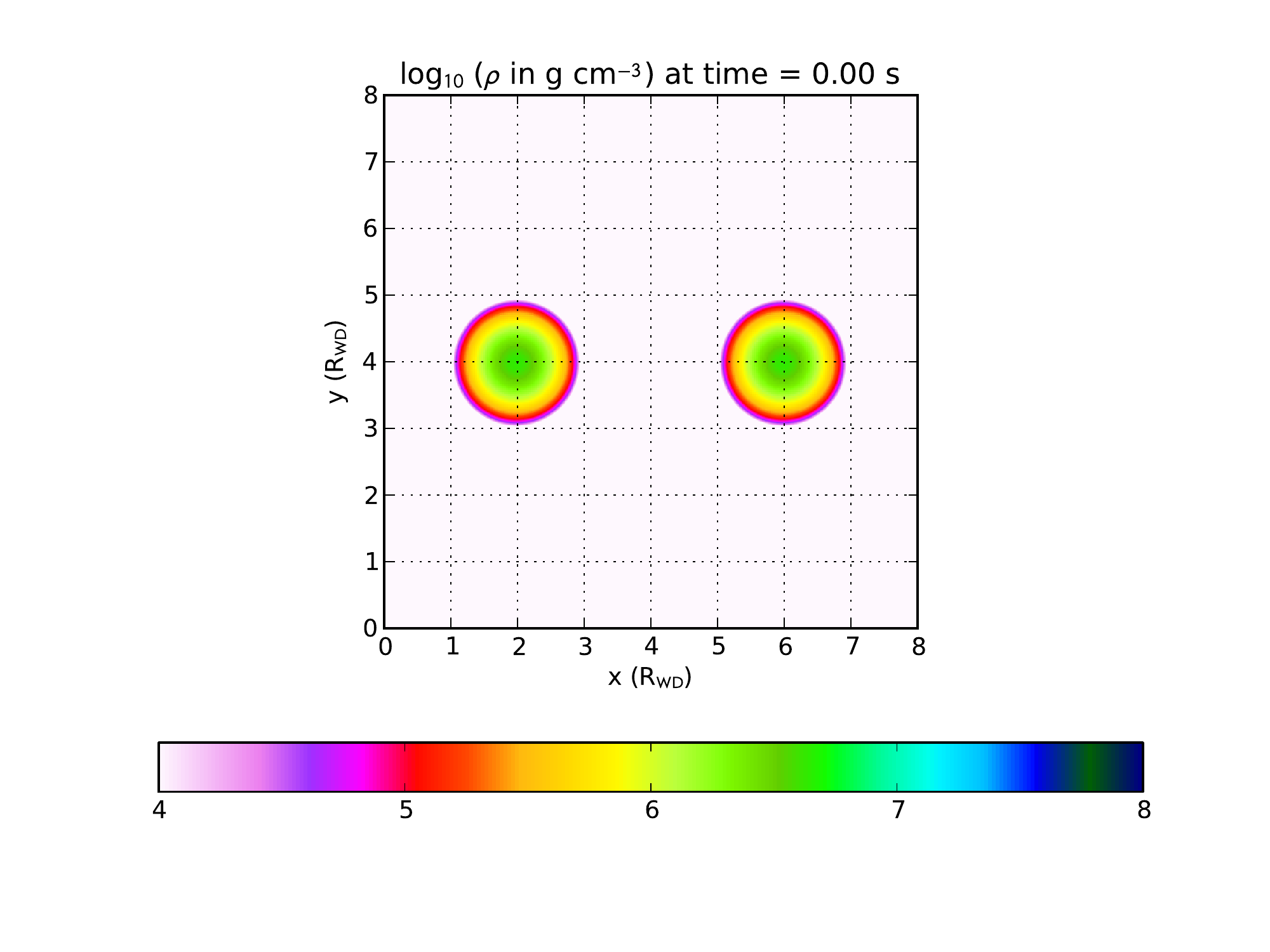}
\end{center}
\caption{
A 2D slice of density through the x-y mid-plane
at t=0.0 s for the 0.64+0.64 M$_{\odot}$ collision. Each
tick mark has a value of one white dwarf radius, which is
8.3$\times$10$^8$ cm. The size of the domain is equal to eight white
dwarf radii, and the white dwarfs are positioned four white dwarf
radii apart from center to center.
}
\label{f.init}
\end{figure}

Our initial 1D white dwarf profiles are calculated assuming
hydrostatic equilibrium and mass conservation\footnote{Code avaliable
from http://cococubed.asu.edu/code\_pages/coldwd.shtml}.  Our initial
white dwarf models use the same equation of state as in FLASH, namely, the
Helmholtz equation of state. We assume a uniform composition of 50\%
\carbon[12] and 50\% \oxygen[16]. We use white dwarfs with masses 0.64
and 0.81 M$_{\odot}$, to match the masses used in
\citet{raskin_2010_aa}, with an isothermal temperature of 10$^{7}$ K.
We map the 1D white dwarf profiles for the density, temperature and
composition onto a 3D rectilinear Cartesian grid. 
 Our computational
domain is a cubic box chosen to be eight times the white dwarf radius
(see Fig.~\ref{f.init}).
The white
dwarfs are initially placed four white dwarf radii apart from center
to center, which is large enough to allow the subsequent evolution to
produce tidal distortions while allowing sufficient numerical resolution in the
central regions.  

The symmetry of head-on collisions between identical, initially
spherical white dwarfs suggests 2D axisymmetric simulations may have
been sufficient.  Our rationale for deploying 3D rectilinear
coordinates is three-fold. First, we want to explicitly show that
FLASH maintains symmetry throughout the collision and subsequent
explosion processes.  Second, we want to compare our results on this
important numerical test case with other existing 3D calculations
(both grid and particle). Imposing axisymmetric conditions would have
complicated these comparisons because we would not know if differences
from existing 3D models were driven by different physics, different
numerics or the imposition of axisymmetry itself.  Third, we
anticipate exploring unequal mass and non-zero impact parameter
collision models with FLASH, both of which violate axisymmetry.  To
better assess the impact of these effects requires an equal mass, zero
impact parameter, 3D benchmark calculation.

We use the free-fall expression for the initial, relative speed of the
two white dwarfs, $v = [2G(M_1 + M_2 )/\Delta r]^{1/2}$, where $M_i$
are the masses of the constituent white dwarfs and $\Delta r$ is the
initial separation of their centers of mass, which for our initial
conditions is 4R$_{\mathrm{WD}}$.  Each white dwarf moves toward the
other white dwarf, one in the positive x-direction and the other in
the negative x-direction, with half of the relative speed.  The
centers of both stars lie on the x-axis, and thus the initial
velocities are purely in the x-direction.

The surrounding ambient medium is set to the
same temperature as the isothermal white dwarfs with a density that is
small (10$^{-4}$ g cm$^{-3}$) compared to the density of the outermost
regions of the white dwarf ($\sim$1-10 g cm$^{-3}$).
Table~\ref{t.initial} lists the initial conditions for each of our
six simulations.

\begin{table}[ht]
\small
\caption{
Initial Conditions for the 3D FLASH models. 
Columns are the run number, white dwarf masses ($M_1$, $M_2$), maximum
level of refinement ($l$), maximum spatial resolution ($R$), domain size
($D$), white dwarf initial velocities ($v_1$, $v_2$), the value of the
timestep limiter ($f$), white dwarf radii ($R_{\mathrm{WD}}$), and
central white dwarf densities ($\rho_{\mathrm{WD}}$).
}
\label{t.initial}
\begin{center}
\begin{tabular}{c c c c c c c c c}
\hline\hline
\# &  $M_1$, $M_2$ & $l$ & $R$ & $D$  & $v_1$, $v_2$ & $f$ & $R_{\mathrm{WD}}$ & $\rho_{\mathrm{WD}}$ \\
&  (M$_{\odot}$) & & (10$^7$ cm) & (10$^9$ cm)  & (10$^8$ cm s$^{-1}$) & & (10$^8$ cm) & (10$^6$ g cm$^{-3}$) \\
\hline
1 & 0.64 & 5 & 5.19 & 6.64 & $\pm$1.59 & 0.2 & 8.30 & 4.51 \\
2 & 0.64 & 6 & 2.59 & 6.64 & $\pm$1.59 & 0.2 & 8.30 & 4.51 \\
3 & 0.64 & 7 & 1.30 & 6.64 & $\pm$1.59 & 0.2 & 8.30 & 4.51 \\
4 & 0.81 & 5 & 4.32 & 5.51 & $\pm$1.97 & 0.2 & 6.88 & 11.2 \\
5 & 0.81 & 6 & 2.16 & 5.51 & $\pm$1.97 & 0.3 & 6.88 & 11.2 \\
6 & 0.81 & 7 & 1.08 & 5.51 & $\pm$1.97 & 0.3 & 6.88 & 11.2 \\
\hline
\end{tabular}
\end{center}
\end{table}

Our FLASH models begin with 1 top-level initial block, where each
block contains 8 cells in each direction $(x, y, z)$. The blocks are
refined or derefined at each time-step based on changes in density and
pressure. For each successive level of refinement, the block size
decreases by a factor of two, creating a nested block structure. 
At maximum refinement, the smallest block size is determined by 
$R = D/(8\times2^{l - 1})$, where $D$ is the domain size in one dimension and $l$ is the maximum
level of refinement as seen in Table~\ref{t.initial}.
At first contact between the white dwarfs, shock waves are sent into
the ambient medium, causing the grid in the ambient medium to rapidly
become maximally refined.  To avoid concentrating resources on these
less interesting regions of the models, we use a derefine procedure at
first contact that sets a radius equal to 1.2 white dwarf radii beyond
which the blocks in the ambient material are forced to be less refined
than the blocks in the collision region.

The nuclear reaction network in FLASH uses constant thermodynamic
conditions over the course of a timestep. However, the Courant limited
hydrodynamic timestep may be so large compared to
the burning timescale that the nuclear energy released in a cell may
exceed the existing specific internal energy. To ensure the hydrodynamics and 
burning remain coupled, as well as to
capture the strong temperature dependence of the nuclear reaction rates,
we limit the timestep as a result of nuclear burning by a factor $f$, 
which constrains the maximum allowable change in specific internal energy. 
The overall timestep is $dt_{n+1} = \min[dt_{\mathrm{hydro}},  
dt_{\mathrm{burn}}]$, where
\begin{equation}
\label{e.enuc2}
dt_{\mathrm{burn}} = dt_{n} \times f \times \frac{u^{i} _{n-1}} {u^{i} _{n}-u^{i} _{n-1}} \text{     ,}
\end{equation}
where the subscript $n$ refers to the timestep number,
$dt_{\mathrm{hydro}}$ is the hydrodynamic timestep,
$dt_{\mathrm{burn}}$ is the burning timestep, and u$^{i}$ is the
specific internal energy of the $i$th cell.
Table~\ref{t.initial} lists the nominal values of $f$ used for our six simulations,
and the effects of using different values of $f$ is discussed in \S\ref{s.convergence}.

\section{Results}
\label{s.results}

\subsection{General Features of the Collision Models}
\label{s.general_features}

Zero impact parameter, or head-on, white dwarf collisions undergo four
distinct phases of evolution. First, the white dwarfs become tidally
distorted as they approach each other. 
For the 0.64+0.64 M$_{\odot}$ case
(hereafter 2$\times$0.64), the velocity gradient across the white
dwarf at first contact ranges from about 3500 km s$^{-1}$ to 5000 km
s$^{-1}$.  Second, a shock wave is produced normal to the x-axis at
first contact. The shock stalls because the speed of infalling
material and the sound speed are comparable.  Third, nuclear burning
is initiated within the stalled shock region.  Finally, the nuclear
energy released unbinds the system, leading eventually to homologous
expansion.

\begin{figure}[tbp]
\begin{center}
\includegraphics[width=1.0\textwidth]{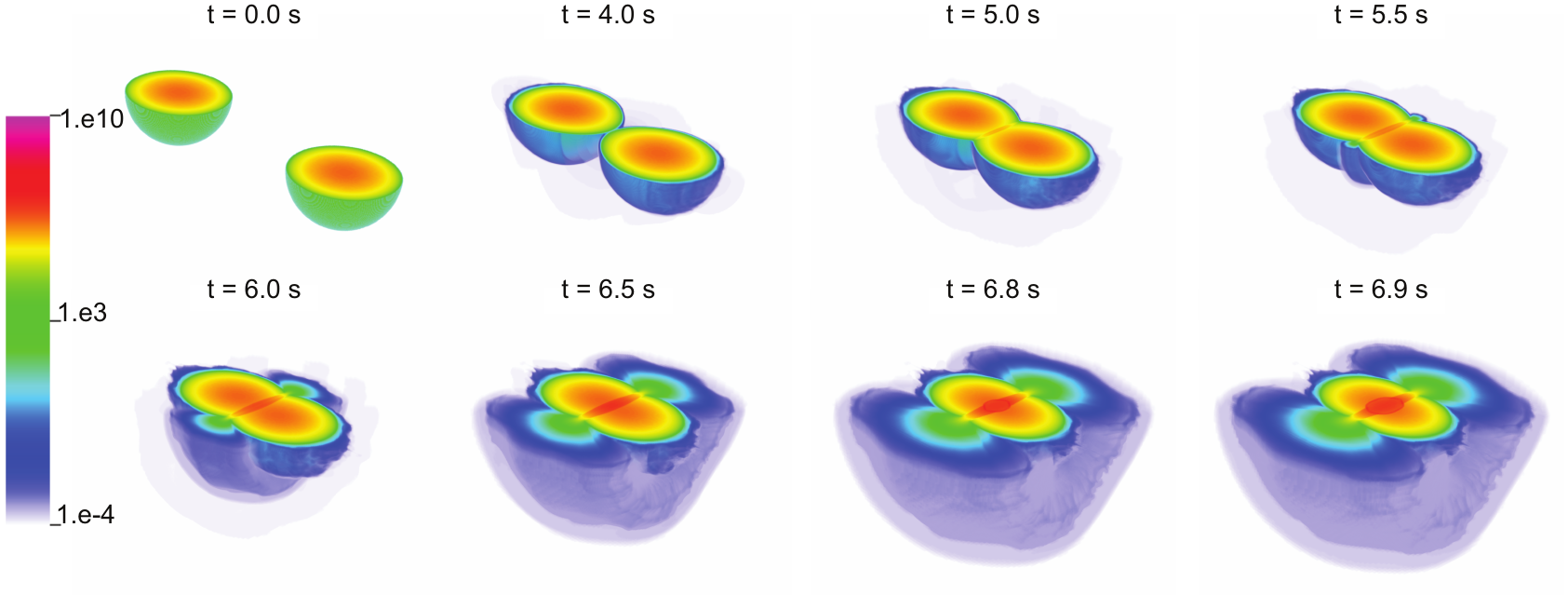}
\end{center}
\caption{3D images cut through the center of the y-axis of the 2$\times$0.64 collision density from first contact at 0.0 s to after ignition at 6.9 s. For scale, each white dwarf has a radius of 8.3$\times$10$^8$ cm. The density colorbar is logarithmic and extends from 10$^{-4}$ to 10$^{10}$ g cm$^{-3}$}.
\label{f.3d}
\end{figure}

An overview of the evolution of the 2$\times$0.64 collision is shown
in Fig.~\ref{f.3d}. The 3D calculation has been sliced through the
 x-z mid-plane to show detail at the center of the
collision. Due to the symmetry of a head-on collision, a cut through the
x-z mid-plane will look identical to a cut through the x-y mid-plane. 
The top panel of the figure represents four times in the
collision from the start of the simulation (t=0.0 s), to first
contact (t=4.0 s), to the formation of the stalled shock region
(t=5.0 s), and finally, to the jettisoning of material
orthogonal to the x-axis (t=5.5 s). 

Given the white dwarf radius and initial velocity
shown in Table~\ref{t.initial} for the 2$\times$0.64 collision, the
time to first contact would be $2R/v$=5.2 s if the initial speed was constant
and the white dwarfs remained spherical. However, the initial speed increases 
due to gravitational acceleration and tidal distortion causes the 
white dwarfs to become elongated along the x-axis. As a result 
the two white dwarfs experience first contact sooner, at about 4.0 s.

The bottom panel represents
the the further progression of the collision from the continued
jettison of material (t=6.0 s), to just before ignition (6.5
s), to just after ignition (t=6.8 s), and finally to the
spread of nuclear burning through the white dwarfs (t=6.9
s). These steps are discussed in further detail below.

\begin{figure}[tbp]
\begin{center}
\includegraphics[width=1.0\textwidth]{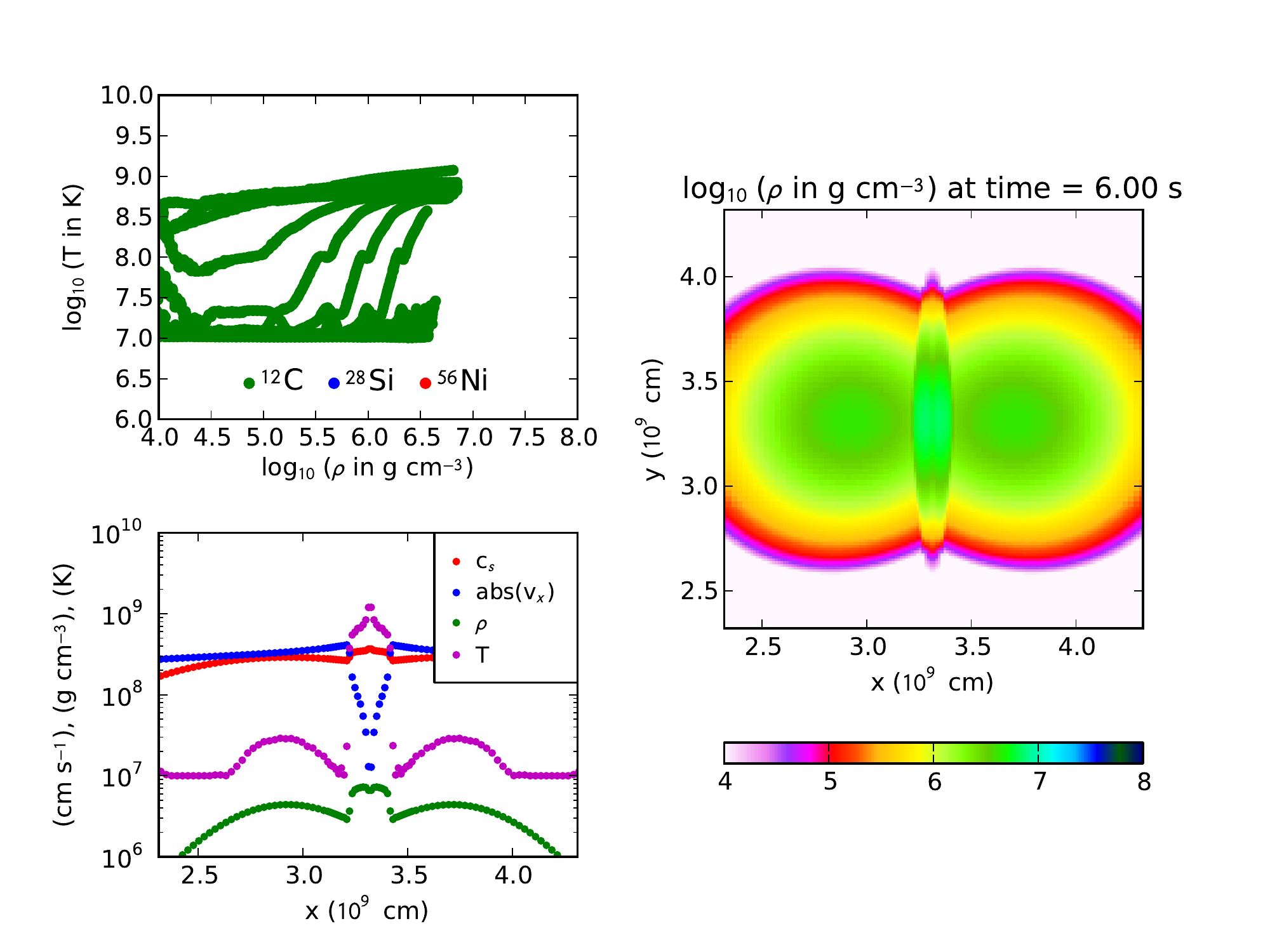}
\end{center}
\caption{
Analysis images of the 2$\times$0.64 collision at t=6.00 s,
after first contact, but before ignition. Top-left: Locations of all cells in the
density-temperature plane. The color of the points represents the
primary composition of the corresponding cell: green for \carbon[12],
blue for \silicon[28], and red for \nickel[56]. The data are binned
into 100 equally spaced bins in logarithmic density and
temperature. Bottom-left: Temperature, x-velocity, density, and sound
speed along the x-axis. Right: A 2D slice of density through the x-y mid-plane.}
\label{f.time0}
\end{figure}

Fig.~\ref{f.time0} shows the thermodynamic, mechanical, and
morphological properties of the 2$\times$0.64 head-on collision
model. At 6.00 s after the beginning of the model, the white dwarfs
are past first contact but have not yet begun runaway nuclear burning.
The right panel shows the mass density profiles of a slice through the
simulation in the x-y plane. In addition to the ambient medium (white
in the figure), there are two distinct regions of density: the
uncollided white dwarf material and the stalled shock region. The
density and temperature are not yet high enough to fuel runaway burning. The lower
left panel of Fig.~\ref{f.time0} shows these quantities as well
as the sound speed and velocity in the x-direction along
a line connecting the centers of the two white dwarfs and 
parallel to the x-axis. The
sound speed is lower than the infall velocity speed, causing the
stalling of the shock region. The temperature profile peaks at
$\approx$10$^9$ K, which is not hot enough to reach the carbon burning
threshold.  The upper left panel of Fig.~\ref{f.time0} shows the state
of the collision in the density-temperature plane. The color of the
points represents the primary composition of the corresponding cell;
green for \carbon[12], blue for \silicon[28], and red for
\nickel[56]. Material with T$\approx$10$^{7}$ K represents the cold
and dense parts of the two stars that have not yet collided.  The
region with 10$^{7}$$<$T$<$10$^{9}$ K and
10$^{4}$$<$$\rho$$<$10$^{6.5}$ g cm$^{-3}$, represents the shocked
material. At this point in the collision, ``tracks'' run from the
lower left to the upper right, representing tori of material
orthogonal to the x-axis at the center of the collision.  In this
case, \silicon[28] and \nickel[56] have not yet been produced, thus
all the cells are primarily composed of \carbon[12].

\begin{figure}[tbp]
\begin{center}
\includegraphics[width=1.0\textwidth]{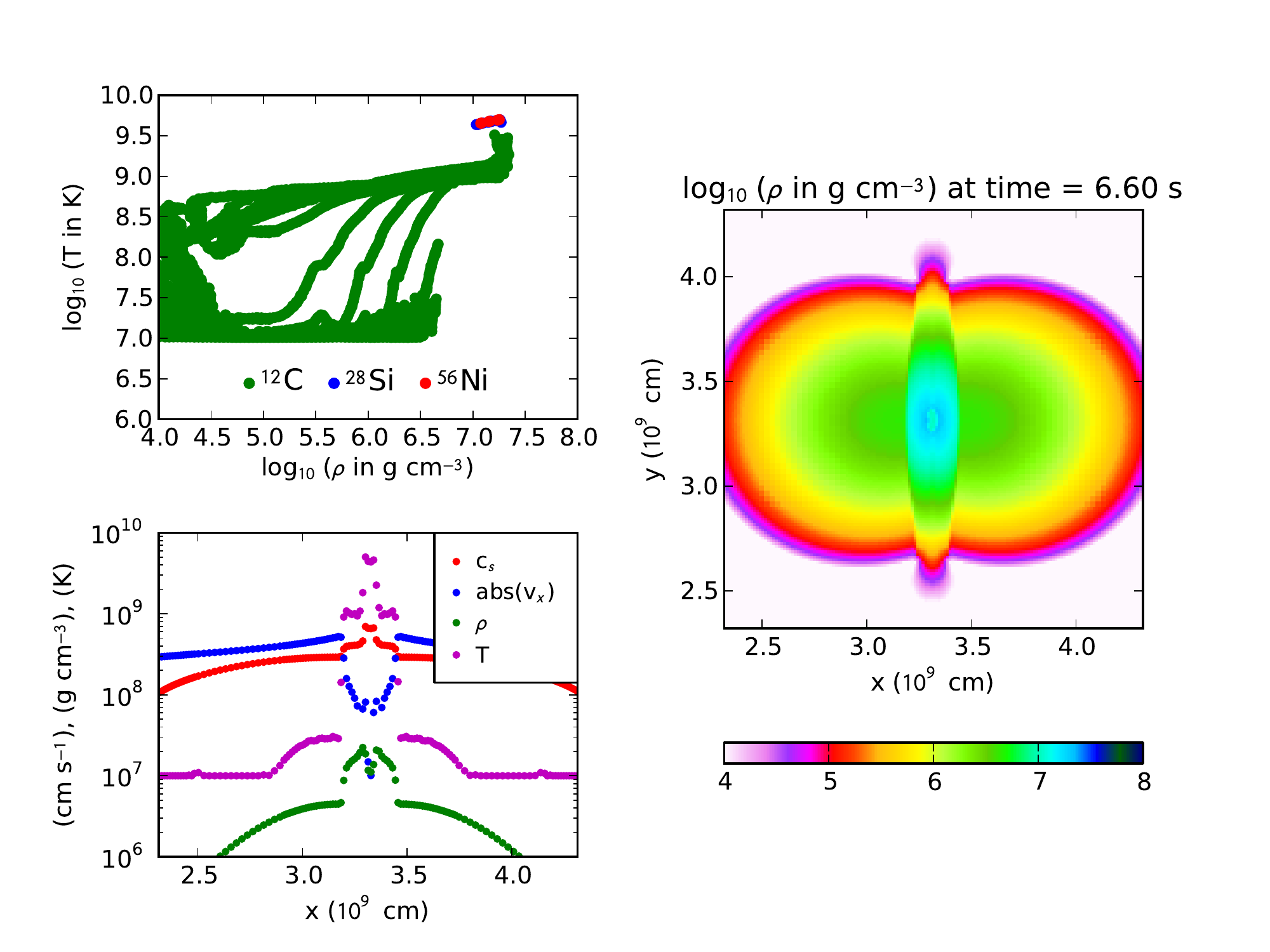}
\end{center}
\caption{
Same format as Fig.~\ref{f.time0}, when the model is at t=6.60 s, right after ignition.}
\label{f.time1}
\end{figure}

Fig.~\ref{f.time1} has the same format as Fig.~\ref{f.time0} at 6.60 s
when runaway nuclear burning has begun.
On the right panel, there are three distinct regions of the 
collision at this point in time: the white dwarf material which has not 
yet experienced the collision; the lenticular, nearly isobaric, stalled shock region;
and the central region where a detonation has begun to propagate.
The detonation front is outlined by the darker colored (higher density)
oval region in Figs.~\ref{f.time1}. Our FLASH simulations do not resolve 
the initiation of the detonation. Instead, at all spatial resolutions 
investigated, the central-most cell in the 2$\times$0.64 head-on collision model 
undergoes runaway carbon burning which begins to propagate a detonation.

In the lower left panel of Fig.~\ref{f.time1}, again, three distinct regions are visible - the
unshocked white dwarfs the stalled shock, and the central-most
detonation region. The temperature in the unshocked white dwarf
material rises smoothly from the initially imposed background
temperature of 1$\times$10$^{7}$ K to $\approx$ 3$\times$10$^7$ K at the
centers of both white dwarfs because of low-amplitude velocity waves
sloshing around the white dwarf interiors. However, 3$\times$10$^7$ K
is well below the carbon burning threshold, does not lift the electron
degeneracy of the material, and does not impact our results.  In the
unshocked region, the infall speed of material is greater than the
local sound speed.  The material behind the stalled shock reaches
temperatures that are sufficient to lift electron degeneracy and are
just below the carbon burning threshold of $\approx$2$\times$10$^{9}$
K. The density in the stalled shock region reaches a peak of
$\approx$2$\times$10$^{7}$ g cm$^{-3}$.  In the innermost region
where a detonation front has traveled $\sim$ 5$\times$10$^{7}$ cm from
the center, the temperature is $\approx$6$\times$10$^{9}$ K and the
density dips to $\approx$1$\times$10$^{7}$ g cm$^{-3}$. 
In the upper left panel of Fig.~\ref{f.time1}, hot, dense
material with T$>$10$^{9}$ K and $\rho$$>$10$^{7}$ g cm$^{-3}$ from
the central regions of the collision are in the upper right corner
where the original carbon material has burned to \silicon[28] and
\nickel[56].

\begin{figure}[tbp]
\begin{center}
\includegraphics[width=1.0\textwidth]{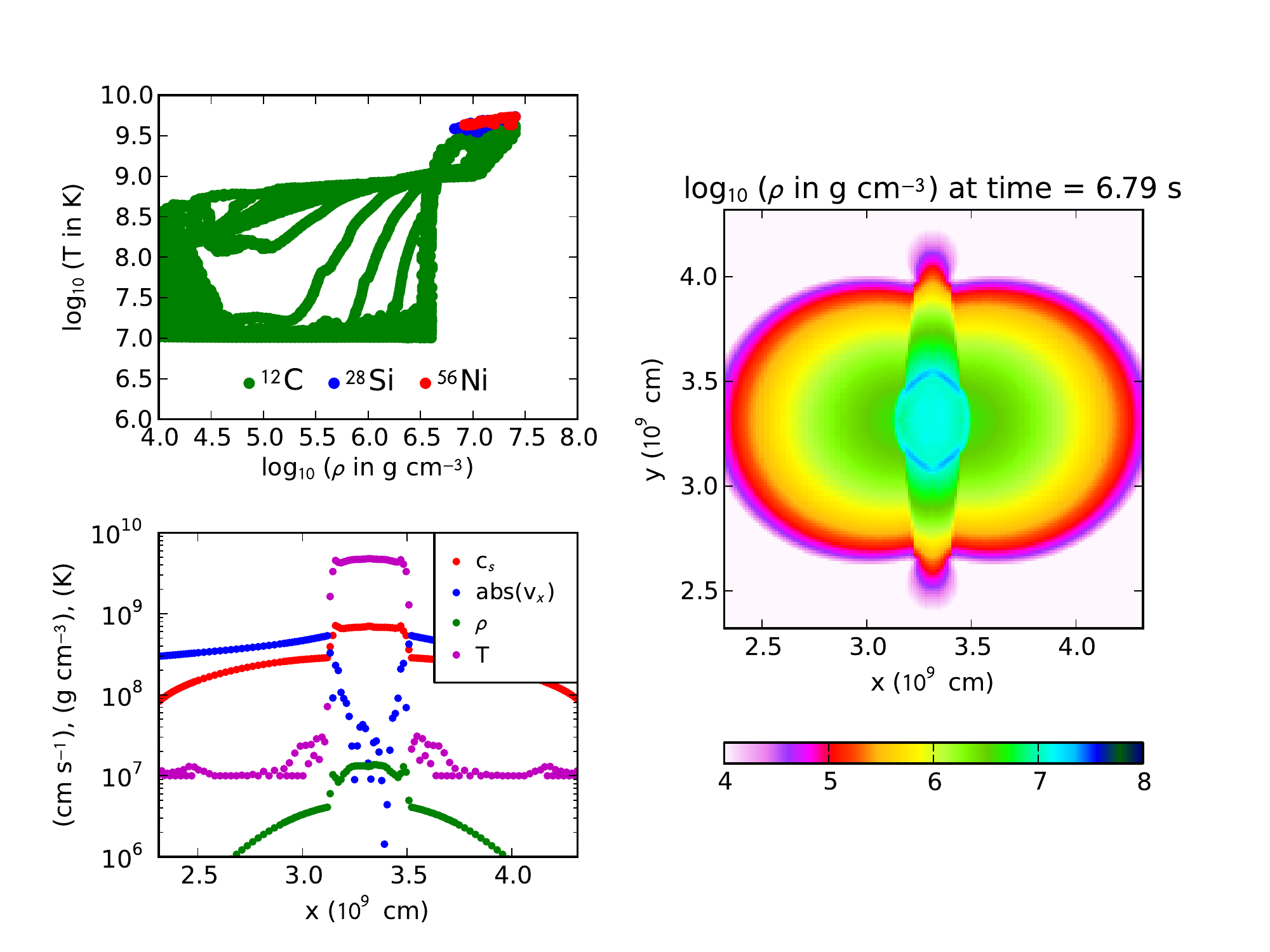}
\end{center}
\caption{Same format as Fig.~\ref{f.time0}, when the model is at t=6.79 s, as the stalled
shock region slightly expands and the densest parts of the white dwarfs begin to enter
the stalled shock region.}
\label{f.time2}
\end{figure}

Fig.~\ref{f.time2} has the same format as Fig.~\ref{f.time0} and the
right panel shows the density profile when the detonation front has
traveled outward from the center and the densest parts of the white
dwarfs are about to enter the stalled shock region.  The upper left
panel indicates that more \carbon[12] material is
present in the high density regime with $\rho$$>$10$^{7}$ g cm$^{-3}$,
and being burned to \silicon[28] and \nickel[56].  The lower left
panel shows that the sound speed in the burned region is comparable with
the speed of the infalling material, and the width of the detonation
has expanded.

\begin{figure}[tbp]
\begin{center}
\includegraphics[width=1.0\textwidth]{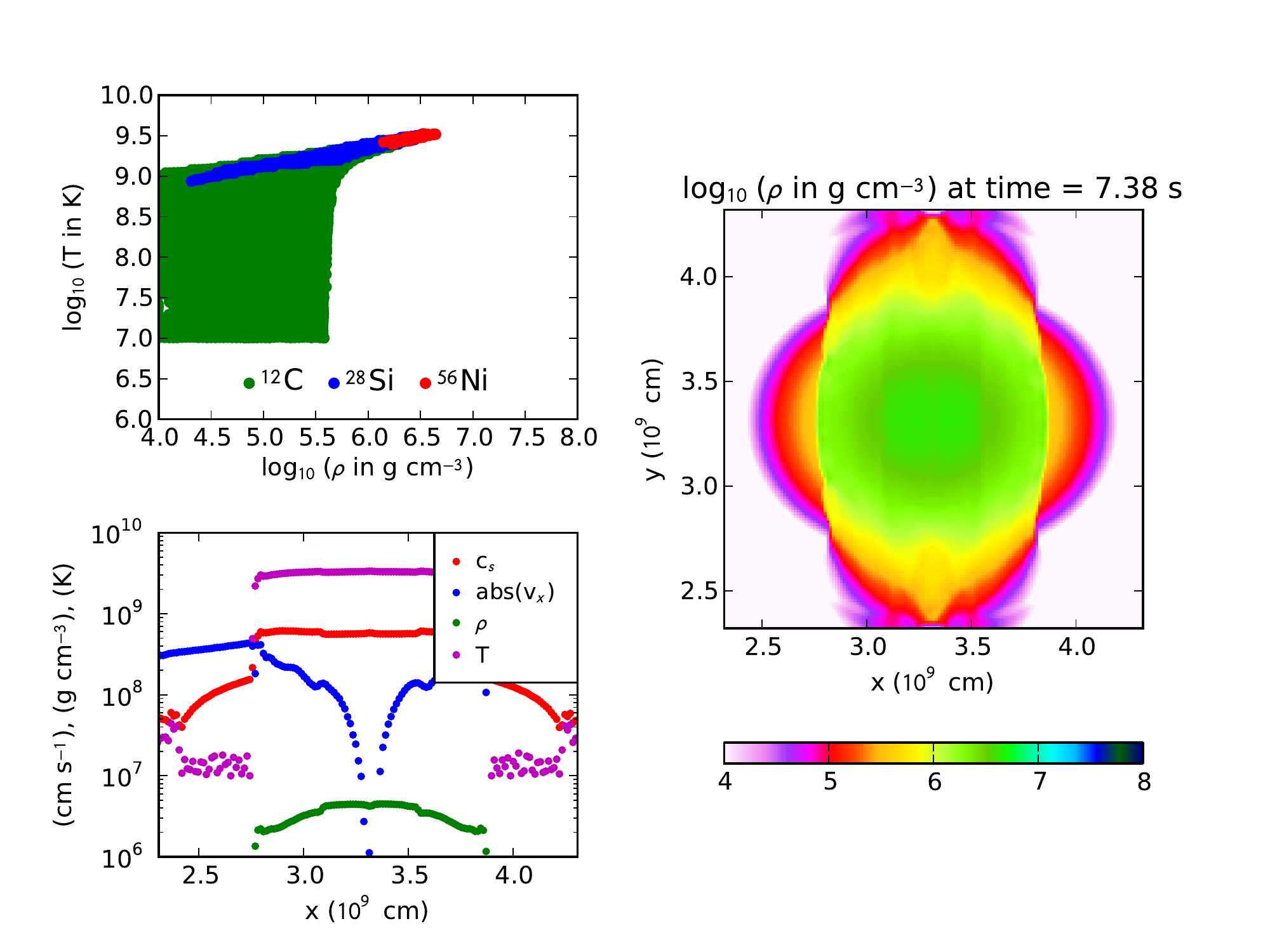}
\end{center}
\caption{Same format as Fig.~\ref{f.time0} at t=7.38 s, just before the system becomes
gravitationally unbound.}
\label{f.time3}
\end{figure}

As additional energy from nuclear burning is added, the double white
dwarf system eventually becomes gravitationally
unbound. Fig.~\ref{f.time3} has the same format as
Fig.~\ref{f.time1}. The right panel shows the density distribution of
the system slightly before the explosion reaches homologous expansion.
The innermost 10$^{9}$ cm reaches a nearly constant temperature of
$\approx$3$\times$10$^{9}$ K with a slowly varying density distribution
that peaks at $\approx$5$\times$10$^{6}$ g cm$^{-3}$.  The
density-temperature plot in the upper left panel indicates larger amounts of
high density, high temperature material with $\rho$$>$10$^{6}$ g
cm$^{-3}$ and T$>$10$^{9}$ K. More material has achieved the
conditions necessary to synthesize \silicon[28] (blue) and \nickel[56]
(red). the lower left panel shows the sound speed is always greater
than the infall speed of the remaining material.

\begin{figure}[tbp]
\begin{center}
\includegraphics[width=1.0\textwidth]{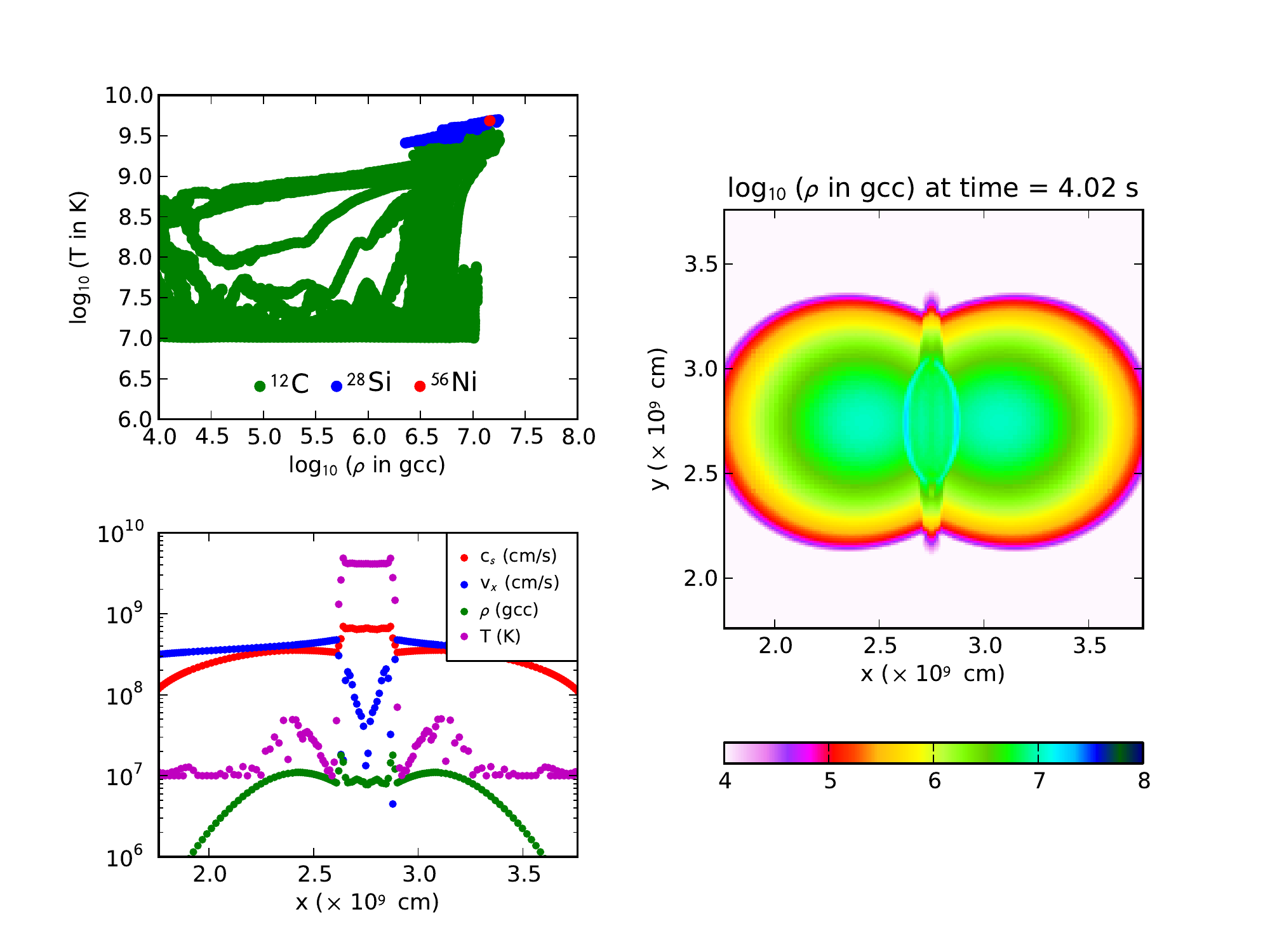}
\end{center}
\caption{Same format as Fig.~\ref{f.time0} at t=4.02 s for the 2$\times$0.81 collision model.}
\label{f.0p81}
\end{figure}

The 0.81-0.81 M$_{\odot}$ (hereafter 2$\times$0.81) collision 
models evolve through a similar set of stages as the 2$\times$0.64 collision models, 
except the larger kinetic energy at impact is sufficient for the initial shock 
to raise the temperature well above the \carbon[12]+\carbon[12] threshold. 
Fig.~\ref{f.0p81} shows that the entire stalled shocked region burns rapidly to 
a state of nuclear statistical equilibrium and achieves a nearly isothermal state.
Central ignition does not occur because the \carbon[12]+\oxygen[16]
material has already been burned to nuclear statistical equilibrium.
We discuss this difference in additional detail in \S\ref{s.explosions}.

Otherwise, the stages of the 2$\times$0.81
collision are very similar to the evolution of the 2$\times$0.64
collision seen above, with the 2$\times$0.81 collision producing a
greater amount of \nickel[56].

\subsection{Numerical Convergence}
\label{s.convergence}

To assess the numerical convergence, we performed the 2$\times$0.64
and 2$\times$0.81 simulations at three different spatial resolutions.
Each increase in spatial resolution is a factor of two more refined in
one dimension, a factor of eight in volume (see
Table~\ref{t.initial}), and takes at least twice as many time steps
depending on the burning timestep. As the spatial resolution
increases, the cells that are burning carbon to heavier elements
become smaller in volume and the timestep decreases, leading to
improved coupling between the hydrodynamics and nuclear burning.

\begin{figure}[tbp]
\begin{center}
\includegraphics[width=0.75\textwidth]{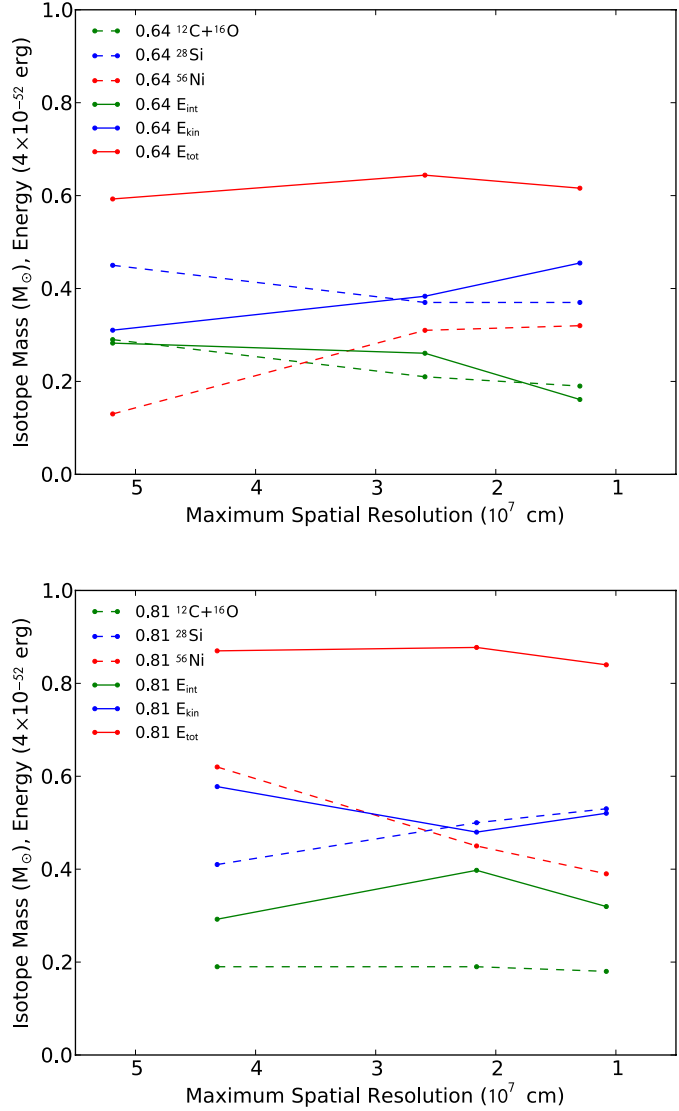}
\end{center}
\caption{
Convergence plot for the 2$\times$0.64 (top) and 2$\times$0.81
(bottom) head-on collisions. Points from left to right correspond to
5-, 6-, and 7-level runs for each collision. The dashed line colors represent
different isotopes, where blue corresponds to \silicon[28], green to
\carbon[12]+\oxygen[16], and red to \nickel[56]. The solid line colors represent 
internal energy (green), kinetic energy (blue), and total energy (red).}
\label{f.converge}
\end{figure}

Table \ref{t.final} lists the ejected masses for each of the six
convergence simulations and Fig.~\ref{f.converge} shows the
convergence behavior of \carbon[12]+\oxygen[16], \silicon[28],
\nickel[56] yields, as well as the internal energy, kinetic energy,
and the total energy at the end of the simulation.  The upper plot in
Fig.~\ref{f.converge} shows that for the 2$\times$0.64 collision the
\nickel[56] mass (dashed red) increases, the \silicon[28] mass (dashed
blue) decreases, and the \carbon[12]+\oxygen[16] (dashed green)
decreases as the spatial resolution increases. The percent change in
\nickel[56] production is 138\% between the 
$R=5.19\times10^{7}$ cm and 
$R=2.59\times10^{7}$ cm models, 
and 3.3\% between the
the $R=2.59\times10^{7}$ cm and 
$R=1.30\times10^{7}$ cm models.  
Although higher resolution models are needed to reach numerical
convergence, the \nickel[56] mass is approaching convergence at 0.32
M$_{\odot}$ (see Table \ref{t.final}). The internal energy (solid
green) at the end of the 2$\times$0.64 collision simulation decreases
with increasing spatial resolution, but the kinetic energy (solid
blue) when the model terminates increases with increasing spatial
resolution.  The net result is that the total energy (solid red) is
nearly constant over the range of resolutions explored.

\begin{table}[ht]
\small
\caption{
Ejected Masses.
}
\label{t.final}
\begin{center}
\begin{tabular}{c c c c c}
\hline\hline
$M_1$, $M_2$ & $R$ & \carbon[12] + \oxygen[16]  & \silicon[28] & \nickel[56] \\
(M$_{\odot}$) & (10$^7$ cm) & (M$_{\odot}$) & (M$_{\odot}$) & (M$_{\odot}$) \\
\hline
0.64 & 5.19 & 0.29 & 0.45 & 0.13 \\
0.64 & 2.59 & 0.21 & 0.37 & 0.31 \\
0.64 & 1.30 & 0.19 & 0.37 & 0.32 \\
0.81 & 4.32 & 0.19 & 0.41 & 0.62 \\
0.81 & 2.16 & 0.19 & 0.50 & 0.45 \\
0.81 & 1.08 & 0.18 & 0.53 & 0.39 \\
\hline
\end{tabular}
\end{center}
\end{table}

The lower panel in Fig.~\ref{f.converge} shows that for the
2$\times$0.81 collision the \nickel[56] mass decreases, the
\silicon[28] mass increases as the spatial resolution increases, and
the \carbon[12]+\oxygen[16] slightly decreases. Although convergence
has not been achieved, the \nickel[56] mass is approaching convergence
at 0.39 M$_{\odot}$ (see Table \ref{t.final}).  We discuss the reason
for the different convergence trends between the 2$\times$0.64 and
2$\times$0.81 cases below.  The internal energy and kinetic energy at
the end of the 2$\times$0.81 collision simulations appears to be
oscillating towards convergence as the spatial resolution is
increased. As a consequence of the internal energy and kinetic energy
being out of phase, the total energy is nearly constant over the range
of resolutions explored.

Although strict numerical convergence has not been achieved with these six
simulations, some trends can be seen. As the
total mass of the binary system increases in zero impact parameter
collisions, the \nickel[56] mass increases, indicating that larger
mass collisions will produce more \nickel[56]. For both mass pairs at
highest resolution, $\approx$0.2 M$_{\odot}$ of unburned
\carbon[12]+\oxygen[16] was ejected.

Higher numerical resolutions are desirable, but prohibitively
expensive for this study, as our most resolved 3D models required at
least 200,000 CPU hours per run.  Simulations with higher spatial
resolution are not possible in the context of the current study
because doubling the grid resolution in a 3D simulation effectively
increases the number of cells by a factor of $\approx 2^3$ and the
number of timesteps by a factor of 2, meaning over an
order-of-magnitude increase in computational time.  Although these
effects can be ameliorated somewhat by adopting more aggressive
derefinement criteria, further restricting the computational domain
size, or relaxing the timestep controller $f$, we expect that
increasing the maximum resolution another factor of two
(6.5$\times$10$^{6}$ cm for the 2$\times$0.64 models and
5.04$\times$10$^{6}$ cm for the 2$\times$0.81 models) would require
$\approx 2$ million CPU hours per run, which is beyond our
capabilities here.

\begin{figure}[htb]
\begin{center}
\includegraphics[width=0.9\textwidth]{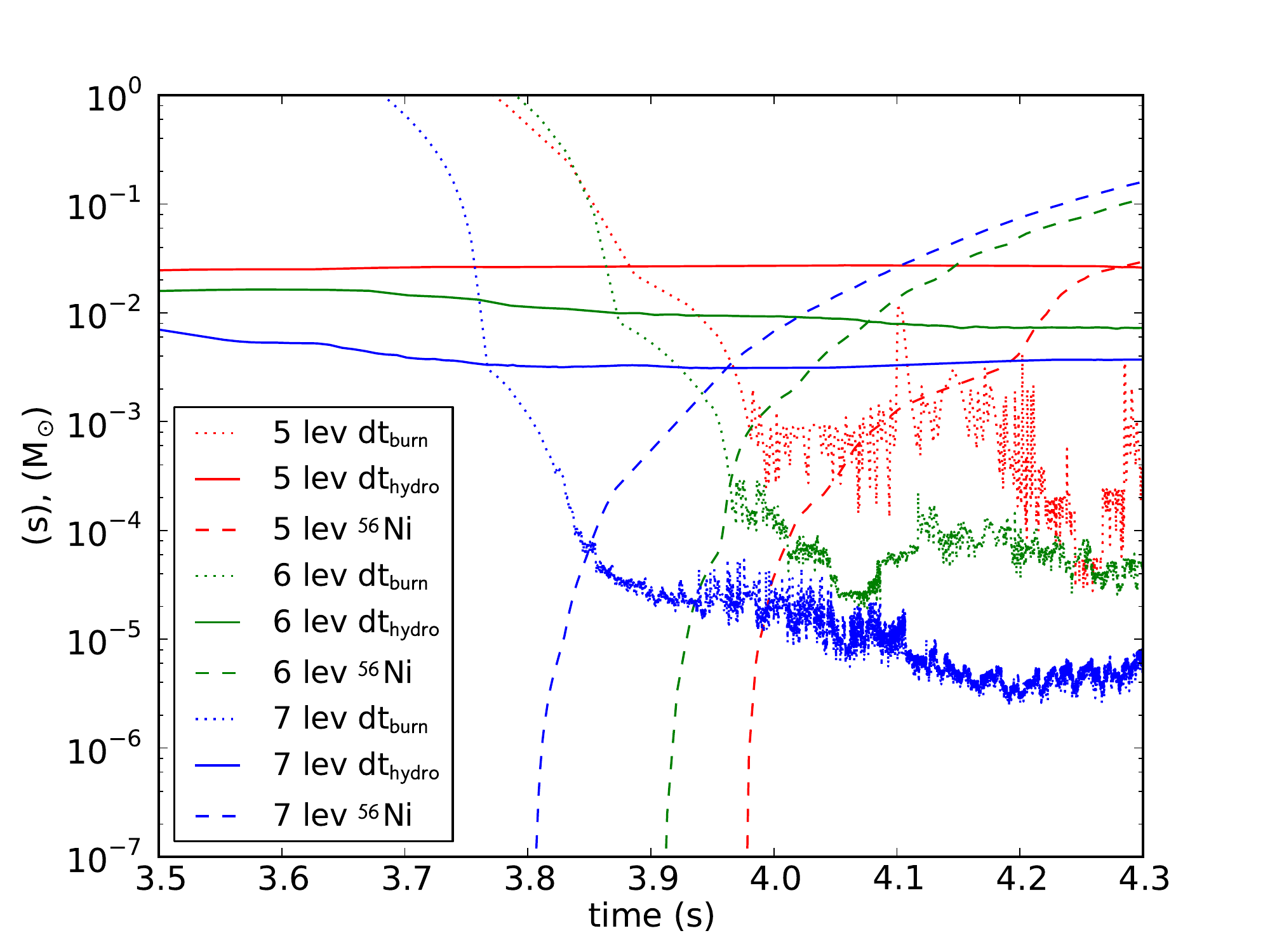}
\end{center}
\caption{
Evolution of the hydrodynamic time step (solid line), burning time
step (dotted line), and \nickel[56] mass (dashed line) for the 5-level
(red), 6-level (green), and 7-level (blue) models of the 2$\times$0.81 collision.
}
\label{f.dt}
\end{figure}

Reducing the timestep limiting factor, $f$, and thereby reducing the
timestep during nuclear burning changes the amount of \nickel[56]
produced.  For example, changing from $f$=0.5 to $f$=0.1 in the
2$\times$0.64 simulation with a spatial resolution of 
$R=2.59\times10^{7}$ cm causes the \nickel[56]
production to increase by approximately 0.1 M$_{\odot}$, a 30\%
change.  Fig.~\ref{f.dt} shows the evolution of the hydrodynamic time
step (solid lines), burning time step (dotted lines), and \nickel[56]
mass (dashed lines) for the 5-level (red), 6-level (green), and
7-level (blue) 2$\times$0.81 collisions. We use $f$=0.2 for the
5-level run and $f$=0.3 for the 6- and 7-level runs to force the
burning timestep to fall below the hydrodynamic timestep during the
\nickel[56] production phase. In all our simulations, we set $f$ such
that dt$_{\mathrm{burn}}\approx$0.01dt$_{\mathrm{hydro}}$ during the
phase of evolution when nuclear burning is significant.  Setting $f$
to smaller values greatly increases the computing time without having a
significant effect on the nucleosynthesis yields.

\begin{figure}[tbp]
\begin{center}
\includegraphics[width=1.0\textwidth]{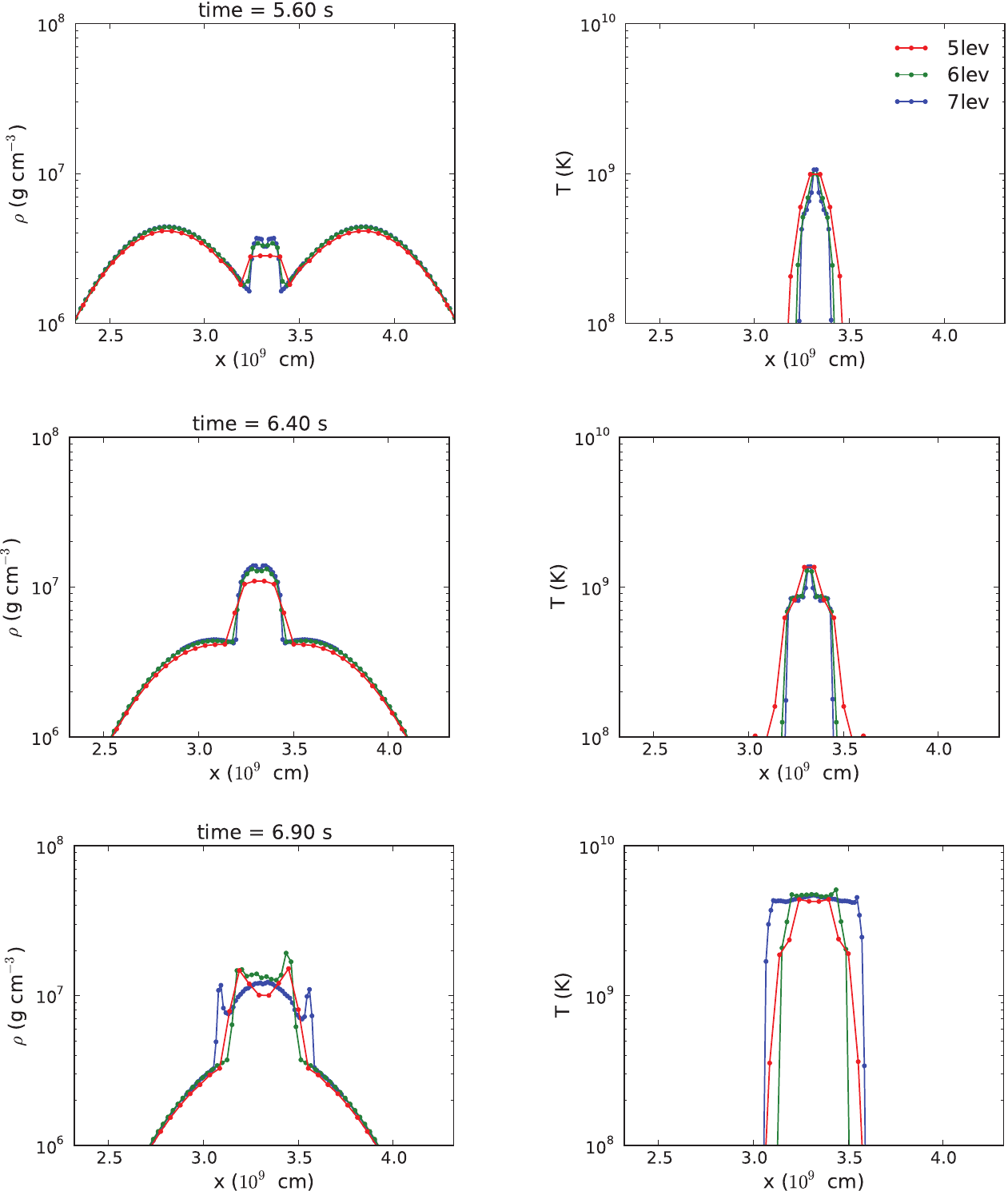}
\caption{Density and temperature profiles along the x-axis 
for the 2$\times$0.64 collision at 5.6 s, 6.4 s, and 6.9 s
 for different levels of refinement.}
\label{f.0p64_movie_stills}
\end{center}
\end{figure}

Fig.~\ref{f.converge} shows that the 2$\times$0.64 collision produces more
\nickel[56] as spatial resolution increases. To understand this
behavior we examine profiles along the x-axis of the density and
temperature for 5-, 6-, and 7-levels of refinement.  The upper panel
of Fig.~\ref{f.0p64_movie_stills} shows the three models with different
spatial resolutions for the 2$\times$0.64 collision at 5.6 s. 
The shocked region is widest in the 5-level model, and
narrower in the 6- and 7-level models.  The density is smaller (just
below $3\times10^6$ g cm$^{-3}$) and nearly constant for the 5-level
model, larger for the 6-level model than the 5-level model
($3.5\times10^6$ g cm$^{-3}$), and slightly larger for the 7-level
model than the 6-level model ($3.6\times10^6$ g cm$^{-3}$).  Both the
6- and 7-level models show a small valley in the central density. The
peak temperature is smaller for the 5-level ($\approx10^9$ K), and
slightly higher for the 6-level and 7-level models (both above $10^9$ K).

At 6.4 s (middle panel in Fig.~\ref{f.0p64_movie_stills}), the density
and temperature profile patterns as described for 5.6 s
generally still hold, but the peak temperature is now the same for all
three levels of refinement ($\approx1.5\times10^9$ K). Detonation
occurs just after this time frame (as seen below). Thus, we 
expect to see more \nickel[56] produced for the 6-level and 7-level models
than for the 5-level model because there is more material in the shocked
region with high density ($>10^7$ g cm$^{-3}$) at the same ignition
temperature. We also expect only a small difference in
\nickel[56] production between the 6- and 7-level models because the
peak density is only slightly larger for the 7-level model and the width of the 
density profile is approximately the same. This explains the pattern in the
the abundance yields with spatial resolution in Fig.~\ref{f.converge}.

At 6.9 s (lower panel in Fig.~\ref{f.0p64_movie_stills}) when the
detonation is underway, the Mach number is larger in the 7-level model
than the 5- and 6-level models because the pre-detonation density is
larger. This causes the 7-level temperature profile to be wider then
the 6-level or the 5-level. That is, the burning front travels farther
for the same amount of time.

\begin{figure}[tbp]
\begin{center}
\includegraphics[width=0.65\textwidth]{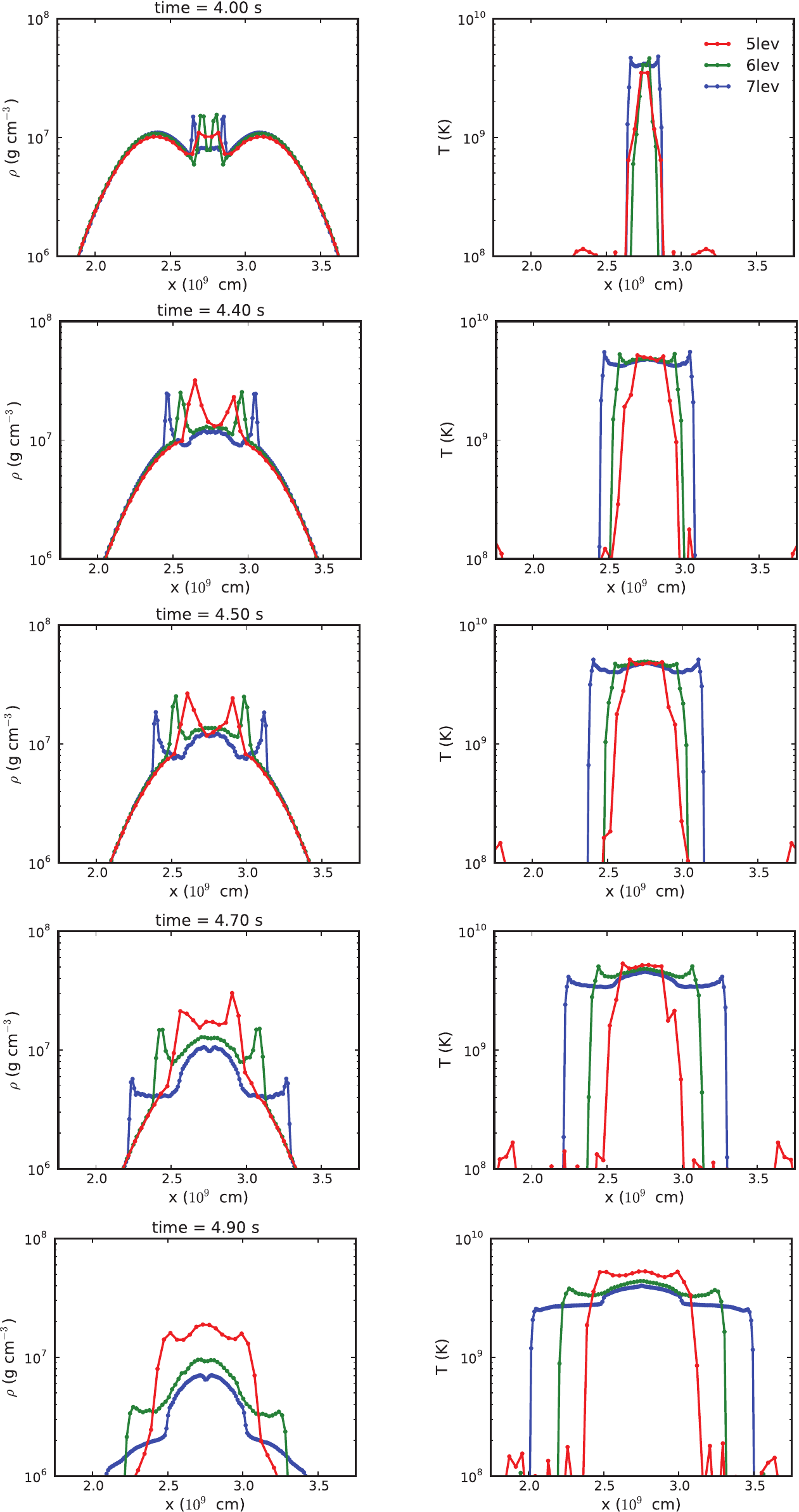}
\caption{Density and temperature profiles along the x-axis 
for the 2$\times$0.81 collision at 4.0 s, 4.4 s, 4.5 s, 4.7 s, and 4.9 s
 for different levels of refinement.} 
\label{f.0p81_movie_stills}
\end{center}
\end{figure}

Unlike the 2$\times$0.64 collision, the 2$\times$0.81 collision
produces less \nickel[56] as level of refinement increases (see
Fig.~\ref{f.converge}). 
The upper panel of Fig.~\ref{f.0p81_movie_stills}
shows the three models with different
spatial resolutions for the 2$\times$0.81 collision at 4.0 s.
The temperatures for all three resolutions are hot ($>3\times10^9$ K), 
indicating the energy generated by burning is large. The shocked region is widest for
the 7-level model and narrowest for the 5-level model with the 6-level model
in-between. The 7-level model has the lowest, and nearly constant, density
($\approx8\times10^6$ g cm$^{-3}$) in the stalled shock region,
and has the largest magnitude spikes in the density (just below $2\times10^7$ g
cm$^{-3}$) at the edges of the stalled shock. The spikes occur because the density of
material is highest just behind the shock front.  The 6-level model has a
slightly larger ($\approx8\times10^6$ g cm$^{-3}$), nearly constant,
density in the middle, and slightly smaller spikes in the density (just below
$2\times10^7$ g cm$^{-3}$) at its edges.  The 5-level model has its 
density spikes (just above $10^7$ g cm$^{-3}$) close enough together that 
a nearly constant density in the middle in barely reached.

The second panel of Fig.~\ref{f.0p81_movie_stills} shows at 4.4 s the
width of the shocked, burning region is larger for all three
resolutions, because the energy generated by burning in the hot
shocked region is sufficient to overcome the standing shock formed
from material moving inwards. That is, the shocked burning region is
growing. The temperature is nearly the same and constant
($\approx5\times10^9$ K) across all three resolutions, but with small
spikes at the edges of each shocked region.  The nearly constant
density in the central region of the 7-level model is still smaller
and wider than the the 6-level model. The 5-level still has its two
spikes near the center, thus a nearly constant central density region
is not reached.

The third panel of Fig.~\ref{f.0p81_movie_stills} shows at 4.5 s the
the 7 level model begins to detonate, but the 6-level and 5-level models
have not yet detonated.  The same patterns in density and temperature 
described for previous time points still hold.
By 4.7 s,  the fourth panel Fig.~\ref{f.0p81_movie_stills} shows the 6-level model begin to
detonate, but the 5 level model has not yet detonated. The width of the
burning region for the 6-level model is about the same width as the 7-level
model when the 7-level model detonated 0.2 s earlier.

At 4.9 s (bottom panel Fig.~\ref{f.0p81_movie_stills}), the 5-level
model is the last to detonate.  The 5-level model has finally reached
a state of nearly constant density in the central region with spikes at the
edges. This nearly constant density of $2\times10^7$ g cm$^{-3}$ is
larger than the nearly constant density reached by either the 6-level
or the 7-level models (both about $10^7$ g cm$^{-3}$), but it has
reached about the same width. Since the 7-level model detonates at the
lowest density (but the same mass since all reach about the same width
before detonating) and soonest in time, the 7-level model should
produce the least amount of \nickel[56]. The 5-level model detonates
at a higher density (and same mass) and latest in time, thus should
produce the most \nickel[56].  This explains the pattern in the the
abundance yields with spatial resolution in Fig.~\ref{f.converge}.

\subsection{Similarities and Differences Between the Explosion Models}
\label{s.explosions}

Whether the explosion is initiated along the edge of the stalled shock
region (as in the 2$\times$0.81 collisions) or in the central regions
of the stalled shock (as in the 2$\times$0.64 collisions) is
controlled by the initial masses of the white dwarfs, as the masses
set the infall speed (escape velocity). The infall speed determines
the strength of the initial shock, and thus the initial post-shock
temperature. In turn, the initial post-shock temperature determines
the amount of initial burning behind the shock, and hence the
temperature profile of the shocked region. Comparing the temperatures
profiles in the shocked region between the 2$\times$0.64 and
2$\times$0.81 collisions, we see that the 2$\times$0.64 temperature
barely reaches $10^9$ K, while for the 2$\times$0.81 temperature is a
hot $\approx5\times10^9$ K over an extended region. The difference in
the temperature between the two model collisions is a direct
consequence of the kinetic energy of the collision (larger kinetic
energy corresponding to larger temperature).

\begin{figure}[tbp]
\begin{center}
\includegraphics[width=0.6\textwidth]{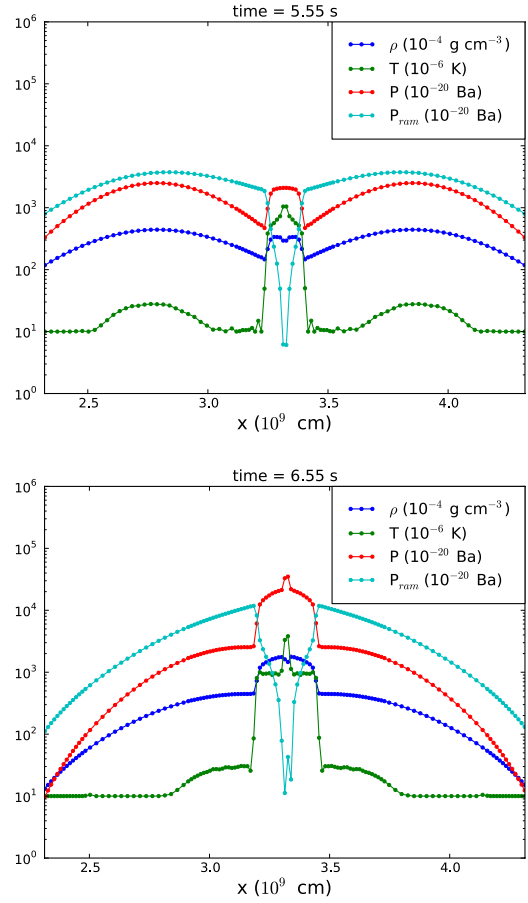}
\end{center}
\caption{Profiles of the density, temperature, pressure, and 
ram pressure along the x-axis for the 2$\times$0.64 collision at 5.55 s and 6.55 s.}
\label{f.0p64_lit}
\end{figure}

The initial shock in the 2$\times$0.64 collision models barely raises
the temperature above the \carbon[12]+\carbon[12] threshold.  As
carbon burning proceeds, the central shocked burning region increases
in temperature.  The top panel of Fig.~\ref{f.0p64_lit}) shows the
system cannot explode yet because the temperature is not hot enough to
overcome the ram pressure from the infalling material, which continues
to increase due to density profile of the white dwarf.  When the peak
of the white dwarf density profiles enters the shocked region does the
central peak undergo thermonuclear runaway, which creates enough
pressure to overcome the now decreasing ram pressure (see bottom panel
of Fig.~\ref{f.0p64_lit}).

\begin{figure}[tbp]
\begin{center}
\includegraphics[width=0.6\textwidth]{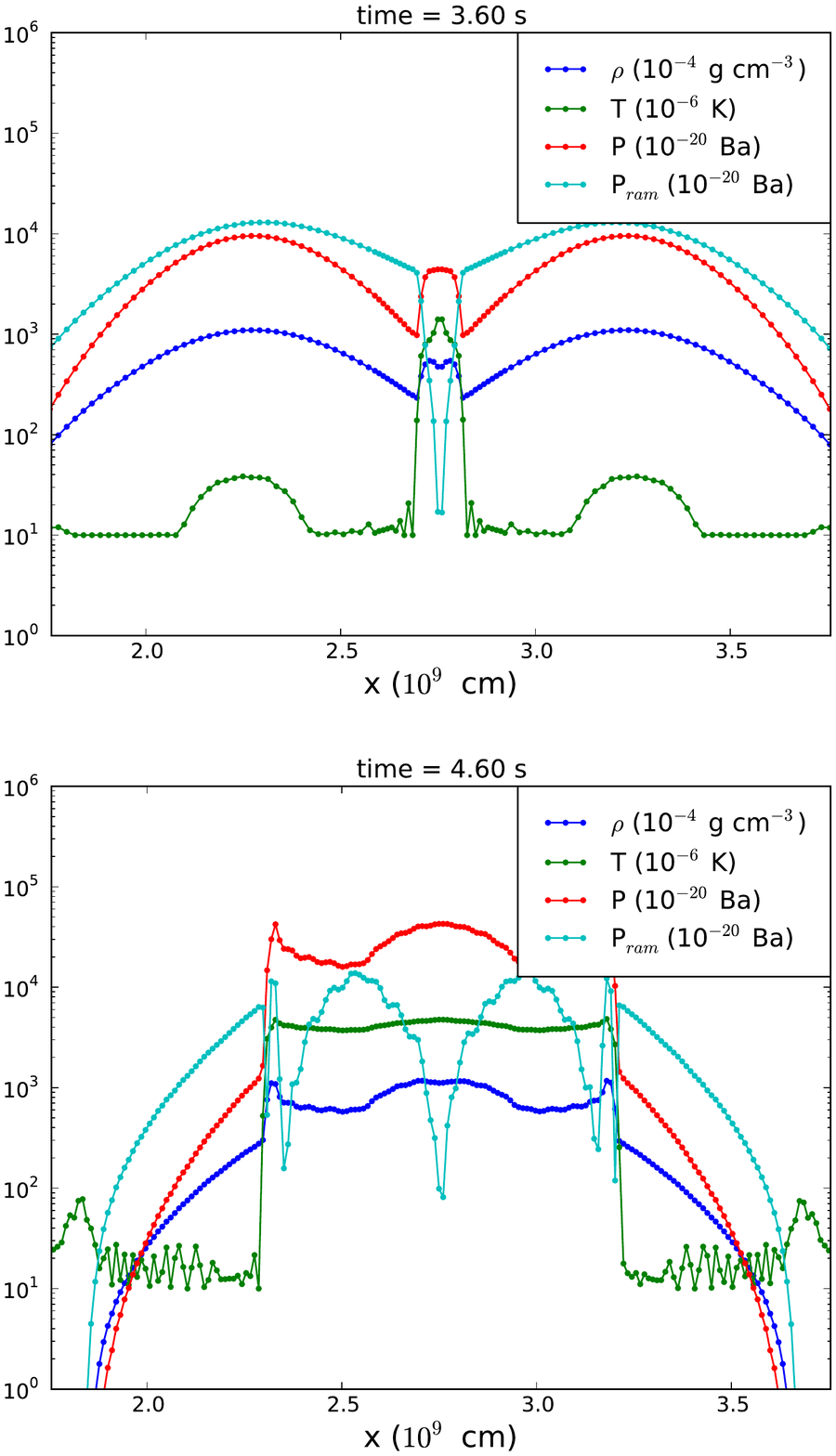}
\end{center}
\caption{Density, temperature, pressure, and ram pressure profiles along 
the x-axis for the 2$\times$0.81 collision at 3.60 s and 4.60 s.}
\label{f.0p81_lit}
\end{figure}

In contrast to the 2$\times$0.64 collision model, the 2$\times$0.81
collision model is energetic enough that the initial shock raises the
temperature well above the \carbon[12]+\carbon[12] threshold. The
entire stalled shocked region burns rapidly to a state of nuclear
statistical equilibrium and achieves a nearly isothermal state.
Central ignition cannot occur because the \carbon[12]+\oxygen[16]
material has already lost nearly all of its energy in the burn to
nuclear statistical equilibrium.  The top panel of
Fig.~\ref{f.0p64_lit} shows, similar to the 2$\times$0.64 collision
models, the 2$\times$0.81 collision model cannot yet explode since the
ram pressure from the infalling material is greater than the pressure
of the hot burned material pushing outwards. When the pressure inside
the hot burned region is larger than the ram pressure does the system
explode, giving the appearance of an edge-lit ignition (see bottom
panel of Fig.~\ref{f.0p64_lit}).

\section{Discussion}
\label{s.discuss}

We have performed the first systematic study of zero impact parameter
collisions between two white dwarfs with an Eulerian grid code
(FLASH).  Our simulations spanned a range of effective spatial
resolutions for collisions between two 0.64 M$_{\odot}$ white dwarfs
and two 0.81 M$_{\odot}$ white dwarfs.  However, even the highest
resolution studies did not achieve strict numerical convergence.

The lack of convergence in the simplest configuration 
(zero impact parameter, equal masses) suggest that 
quantitative predictions of the ejected elemental abundances 
that are generated by binary white dwarf collision and merger
simulations should be viewed with caution.
However, the convergence trends do allow some patterns to be discerned.

We found the
2$\times$0.64 collision model head-on collision model produces 0.32
M$_{\odot}$ of \nickel[56], 0.38 M$_{\odot}$ of \silicon[28], and 0.2
M$_{\odot}$ of unburned \carbon[12]+\oxygen[16].
\citet{rosswog_2009_aa} included one FLASH based model of a zero
impact parameter collision of two 0.60 M$_{\odot}$ white dwarfs in
their study. They reported a \nickel[56] mass of 0.16 M$_{\odot}$,
about one-half of what we find. While \citet{rosswog_2009_aa} used
slightly less massive white dwarfs than our study, both sets of FLASH
simulations used the same equation of state.  The FLASH model of
\citet{rosswog_2009_aa} achieved about a factor of 2.6 smaller spatial
resolution than our study $R=4.9\times10^{6}$
cm versus $R=1.3\times10^{7}$ cm), due to their
evolving one white dwarf and deploying a mirrored gravitational
potential.  This difference in the maximum spatial resolution could
account for the different \nickel[56] masses between the two
calculations, although the convergence trend shown in the upper panel
Fig.~\ref{f.converge} suggests spatial resolution might not be the
only reason for the difference.  Another potential reason for the
difference in the \nickel[56] masses is the choice of the timestep,
and thus the coupling between the operator split processes of
hydrodynamics and the nuclear burning.  In all our simulations, we
limited the timestep to $\approx$0.01 of the Courant limited
hydrodynamic timestep during the nuclear burning phases (see
Fig.~\ref{f.dt}).  We found changing the allowed timestep can change
the \nickel[56] mass produced by 30\% - 40\%.

We find our FLASH-based, zero impact parameter, collision models
systematically produce less \nickel[56] and more silicon-group
elements than collisions models calculated with SNSPH by
\citet{raskin_2010_aa}. This difference between particle and grid
based codes was first found by \cite{rosswog_2009_aa}, who suggested
that differences in nuclear reaction networks or advection effects
could be responsible for the different yields.  While our FLASH models
used the same equation of state and nuclear reaction network as
\citet{raskin_2010_aa}, and we checked the same output values were
returned for the same input values, a detailed investigation of the
differences between our FLASH model results and the
\citet{raskin_2010_aa} results with SNSPH are beyond the scope of this
paper.

Red and dim SNIa such as SN~1991bg
\citep{leibundgut_1993_aa,turatto_1996_aa,hachinger_2009_aa} SN~1992K
\citep{hamuy_1994_aa}, SN~1999by \citep{garnavich_2004_aa}, and
SN~2005bl \citep{taubenberger_2008_aa} are characterized by
M$_{\mathrm{V}}\approx$-17.  The light curves of underluminous SNIa
decline even more rapidly than expected from a linear luminosity to
decline-rate relation among normal SNIa \citep{phillips_1999_aa,
  taubenberger_2008_aa,blondin_2012_aa}.  Spectroscopically, 91bg-like
SNIa show low line velocities around B-magnitude maximum
\citep{filippenko_1992_aa} and clear spectral signatures of Ti-II,
indicating lower ionization \citep{mazzali_1997_aa}.  Taken together,
these properties together are consistent with $\sim$0.1 M$_{\odot}$ of
newly synthesized \nickel[56].  Our FLASH models suggest 2$\times$0.64
M$_{\odot}$ and 2$\times$0.81 M$_{\odot}$ head-on collision models
produce \nickel[56] masses below that needed for normal SNIa, but are
within a range consistent with observations of underlumnous SNIa. In
addition, either mass pairing produces $\sim$ 0.2 M$_{\odot}$ of
unburned C+O, which may be a unique signature of mergers and
collisions between white dwarfs.

Future studies should include a survey of non-zero impact parameter
white dwarf collisions with FLASH, an exploration of unequal mass
collisions, and an investigation why Lagrangian particles codes and
Eulerian grid codes continue to find about a factor of two difference
in the mass of \nickel[56] ejected.  The zero impact parameter is
insightful as an upper limit on \nickel[56] production, but a non-zero
impact parameter study will likely give a range of \nickel[56] yields
for different collision configurations. An exploration of unequal mass
collisions could provide a broader physical parameter space and allow
an improved quantitative description of how SNIa luminosity scales
with mass pairings.

\section{Acknowledgments}
This work was supported by the National Science Foundation under grant
AST 08-06720 and through the Joint Institute for Nuclear Astrophysics
(JINA) under grant PHY 02-16783. All simulations were conducted with
Arizona State University Advanced Computing Center and Extreme Science
and Engineering Discovery Environment (XSEDE) compute resources. FLASH
was in part developed by the DOE-supported ASC/Alliances Center for
Astrophysical Thermonuclear Flashes at the University of Chicago. 
Wendy Hawley thanks Brandon Mechtley for his invaluable computing assistance.

\bibliographystyle{apj}                                                                  

\bibliography{fxt_master}

\end{document}